\documentclass[aip, amsmath, amssymb, preprint]{revtex4-1}
\usepackage{graphicx}
\usepackage{dcolumn}

\newcommand{\RE}{\mathrm{Re}\,}
\newcommand{\IM}{\mathrm{Im}\,}

\begin{document}

\title{Plasmonic enhancement of the infrared radiation absorption in ultrathin InSb layer}

\author{Yurii M. Lyaschuk}
\email{yulashchuk@gmail.com}
\affiliation{Department of kinetic phenomena and polaritonics, Institute of Semiconductor Physics of NAS of Ukraine, Kyiv, Ukraine}

\author{Vadym V. Korotyeyev}
\email{koroteev@ukr.net}
\affiliation{Department of Theoretical Physics, Institute of Semiconductor Physics of NAS of Ukraine, Kyiv, Ukraine}

\author{Viacheslav A. Kochelap }
\email{kochelap@ukr.net}
\affiliation{Department of Theoretical Physics, Institute of Semiconductor Physics of NAS of Ukraine, Kyiv, Ukraine}

\author{Oleksandr O. Raichev}

\affiliation{Department of Theoretical Physics, Institute of Semiconductor Physics of NAS of Ukraine, Kyiv, Ukraine}

\begin{abstract}
Indium antimonide (InSb) is a fundamental material for infrared radiation detectors based on interband transitions. Its narrow bandgap enables detection of infrared radiation within the $3-5\ \mu m$ atmospheric window, while its high quantum efficiency ensures excellent sensitivity in InSb-based detectors. We propose a plasmonic structure that significantly enhances infrared absorption in an ultrathin InSb film. The resonant characteristics of this plasmonic enhancement effect could serve as a foundation for developing highly sensitive multi-color detectors.
\end{abstract}
\maketitle

\section{Introduction}

The application of infrared detection ranging from astronomy, medical, and military applications to environmental applications such as atmospheric gas sensing \cite{InfraredDetectionBook, spectroscopy1999infrared, Remote2021}. There are two atmospheric windows for infrared radiation: middle wavelength IR(MWIR), which corresponds to the wavelength range 3-5 $\mu m$, and long wavelength IR (LWIR), which corresponds to the range 8-14 $\mu m$ \cite{rogalski2003infrared}.
Indium antimonide is a well-established material used for the fabrication of MWIR infrared detectors \cite{avery1957insbdetector, HULME1962211, rogalski2012history}.
InSb has the narrowest bandgap within $A_{III}B_{IV}$ compounds, which varies from 0.17 eV at room temperature to 0.235 eV at cryogenic cooling conditions \cite{razeghi2003overview}.
The temperature dependence of the bandgap width can be approximated by a simple relation \cite{littler1985temperature_dependence}:
\begin{equation}\label{InSbgap}
   E_g(T) = E_g(0) - \frac{\delta T^2}{\beta+T}
\end{equation}
where $T$ is absolute temperature, $E_g(0) = 0.235\ eV$, $\delta = 6.0\times10^{-4}\ eV/K$, $\beta = 500\ K$.

Although more advanced materials for infrared detection, such as HgCdTe, have been developed, InSb photodetectors remain popular due to their simpler and more cost-effective fabrication process.
InSb photodiodes can be used to build a Focal Plane Array (FPA), which has excellent quantum efficiency, uniformity, and high
pixel operability \cite{Klipstein2013InSbIsrael}. Due to their excellent characteristics, InSb-based FPA was used in sensors of the Spitzer cosmic telescope \cite{SpitzerSpaceTelescope}.
It is worth noting that the use of HgCdTe introduces inherent risks of heavy metal pollution at the manufacturing sites and at the end of the lifecycle without proper recycling. In contrast to it, InSb detectors are more environmentally friendly.
However, standard InSb detectors suffer from a big generation-recombination noise related to the defect levels \cite{rogalski2012history}.
To reduce noise and improve sensitivity InSb-based photodiodes require liquid nitrogen cooling, and therefore, the necessity of expensive and bulky cooling equipment limits their usage.

Therefore, improved fabrication techniques or geometrical engineering are needed to improve the characteristics of the InSb diodes.
For instance, an improved barrier material composition, MBE usage, and the p-n-junction optimization allow both noise and parasitic currents reduction.
Reduction of the noise can help relax the cooling requirements or even allow the creation of room-temperature InSb-based detectors \cite{RoomTInSBUeno_2013}.

One of the well-established methods to improve photodetection is utilizing surface plasmon resonance. It was used to design various detectors with an improved performance of  InAs detectors \cite{zhou2023SPP_InAs}, quantum well detectors \cite{Wu2010QWellEnhanced}, solar cells \cite{Pillary2007SPPSolar}, and HgCdTe MWIR photodetectors \cite{gratingHgCdTeSPP}.
Another approach is based on the excitation of local plasmons, grating 1D-plasmons \cite{Ploog1991PRL}.

We propose a plasmonic structure that can help increase the absorption of infrared radiation and allows the use of ultrathin films of InSb instead of bulk crystals. The reduced thickness of the active layer in ultrathin InSb films can result in a stronger built-in electric field within the p-n junction. This stronger electric field promotes more effective separation and collection of photogenerated electron-hole pairs, reducing the probability of recombination events and consequently lowering recombination noise.
 Moreover, the fabrication process of ultrathin films (e.g., molecular beam epitaxy or metal-organic chemical vapor deposition) can provide better control over the material's defect and impurity levels compared to bulk crystals, potentially reducing the number of recombination centers and the associated recombination noise.
 The proposed structure is composed of an Indium Antimonide (InSb) substrate with a gold grating on top Fig.~\ref{Fig1}(a).

\section{Methods}

Calculation and analysis of the far-field and near-field properties were carried out using well-established numerical methods of computational electrodynamics. The main method was improved RCWA based on the enhanced transmittance matrix approach\cite{MoharamRCWA81, Moharam95, Lalanne96}. The method was implemented using free and open-source Python numerical libraries NumPy and SciPy \cite{harris2020array,2020SciPy-NMeth}. Additionally, for the limiting case of a thin grating, the integral equation method was applied. The main advantage of this method is guaranteed convergence \cite{Nosich2013Noble}. Moreover, it has fast convergence rates, which enable a time-efficient numerical investigation for the different structures that include thin metal or graphene-strip gratings \cite{Nosich2013Noble, Nosich2013Graphen} or quantum-wire gratings \cite{MikhPRB98}. But for the general case of an arbitrary thickness grating, this method can not be applied.  Therefore, the method of the integral equation was used for the initial investigation of the different structure configurations and material parameters, but eventually, these results were validated using the RCWA method.

 A frequency-dependent dielectric function of InSb has been described by a simple expression that was obtained from the microscopic theory of light absorption \cite{adachi1987dielectric, miyazaki1991model_temperature}.

\begin{equation}\label{ImInSbEpsilon}
  \IM \varepsilon(\omega) = \left\{
   \begin{array}{c}
   \displaystyle{\frac{\alpha}{(\hbar\omega)^2} (\hbar\omega - E_g)^{1/2}},\ \hbar\omega>E_g \\
  0, \hbar\omega<E_g
  \end{array}
   \right .
\end{equation}

\begin{equation}\label{ReInSbEpsilon}
  \RE \varepsilon(\omega) = \left\{
   \begin{array}{c}
   \displaystyle{\frac{\alpha}{(\hbar\omega)^2}\left(2-(\hbar\omega + E_g)^{1/2}) + (\hbar\omega - E_g)^{1/2}\right)},\ \hbar\omega>E_g \\
   \displaystyle{\frac{\alpha}{(\hbar\omega)^2}\left(2-(\hbar\omega + E_g)^{1/2}\right)},\ \hbar\omega<E_g,
  \end{array}
   \right .
\end{equation}
where $E_g$ is the InSb bandgap width.
This expression was used for the structure spectra and the partial absorption calculation.
For the calculation of the near field, it was used the method that helps to alleviate the influence of Gibbs phenomena~\cite{lalanne1998computation, Weismann_2015}.

\section{Results and Discussion}

\begin{figure}[h]
  \centering
  \includegraphics[width=0.5\textwidth]{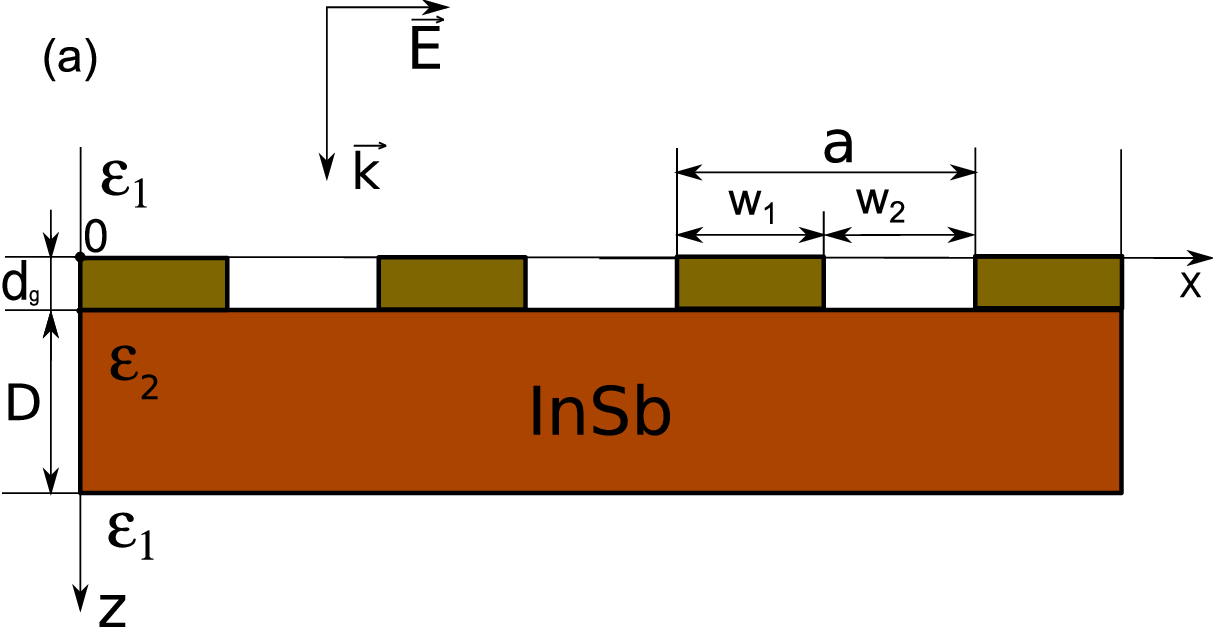}
  \includegraphics[width=0.45\textwidth]{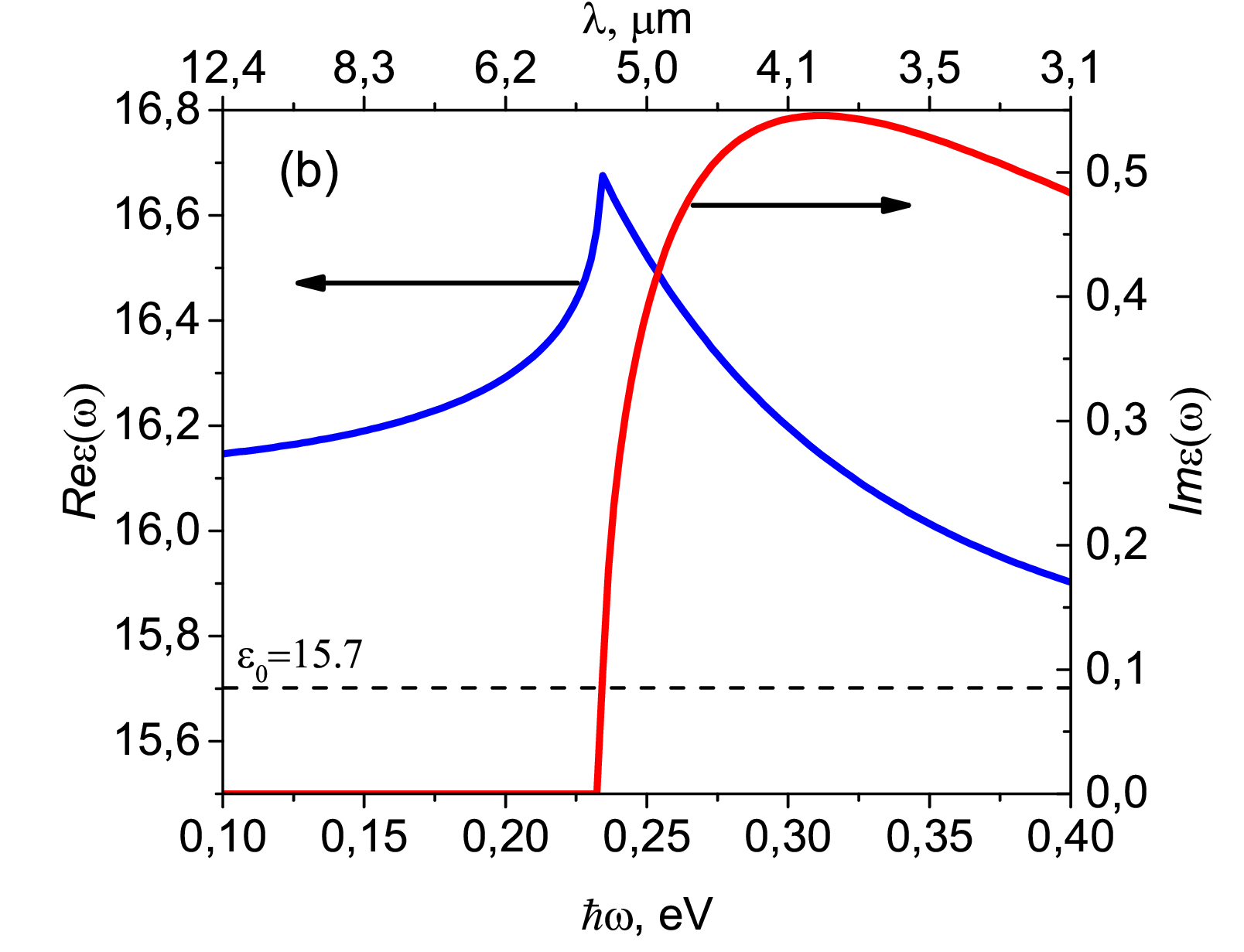}\\
  \caption{Panel a: the schematic view of the proposed structure, Panel b: the dependence of the real and imaginary parts of InSb dielectric permittivity near the fundamental edge of absorption.}\label{Fig1}
\end{figure}

The structure in Fig.~\ref{Fig1} is illuminated by a TM-polarized electromagnetic wave, which induces instantaneous dipole charges and currents. These charges and currents give rise to the local electromagnetic field (or the near field). Generally, such redistribution of the incident wave power does not significantly affect the InSb substrate absorption: in the areas where the grating near field is much stronger, InSb absorption increases, but in areas where the near field is weaker, the opposite effect occurs. As a result, the net effect is nearly the same as in the case without a grating. However, during the plasmonic resonance, the situation changes drastically. The intensity of the plasmon-related near field increases by an order of magnitude. Consequently, the average absorption increases.

\subsection{Results}
\begin{figure}[h]
  \centering
  \includegraphics[width=0.8\textwidth]{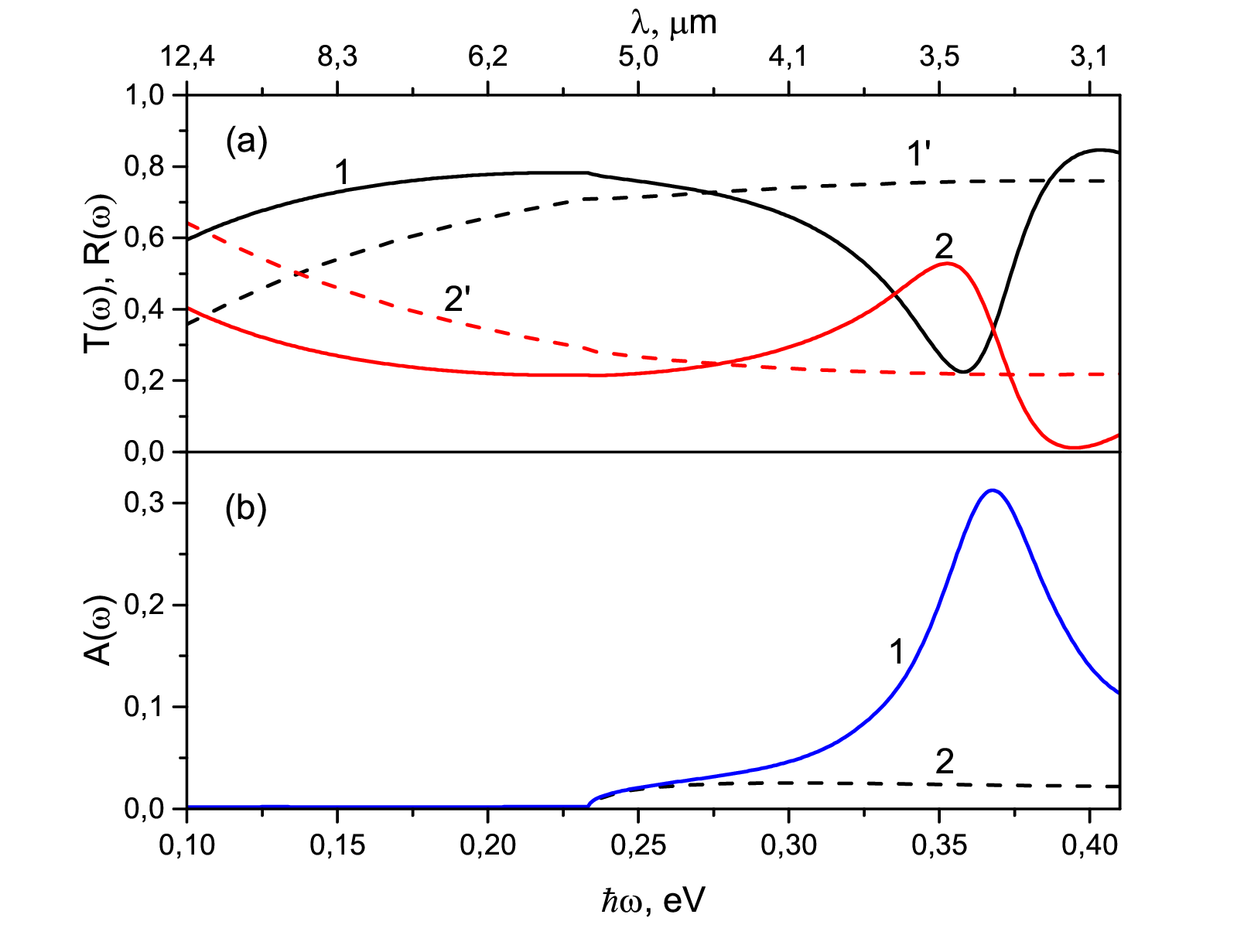}\\
  \caption{The spectra of transmission, reflection (a), and absorption (b) of the plasmonic structure in comparison to the spectra of a bare substrate. Panel a: 1,2 are the reflection and absorption spectra of the plasmonic structure 1', 2' are the corresponding ones for the bare substrate, and panel b: 1, 2 are the absorption spectra of the grating structure and bare substrate, respectively. The thickness of the substrate  $D = 300$ nm. The parameters of InSb are: $\alpha = 0.19\ eV^{3/2}$, the bandgap $E_g = 0.23\ eV$, the static dielectric permitivitty  $\varepsilon_0 = 15.7$. The grating parameters are: the grating period $a = 1.5\ \mu m$, the grating depth $d = 100\ nm$, the metal conductivity
  $\sigma_0 = 3.7\times10^{17}\ s^{-1}$,
  the momentum relaxation time $\tau = 25\ fs$
  the strip width $w_1 = 0.36\ \mu m$ and the opening width $w_2 = 1.14\ \mu m$.}\label{Fig2}.

\end{figure}
Fig.~\ref{Fig2} demonstrates a comparison of transmission $T(\omega)$, reflection $R(\omega)$, and absorption $A(\omega) = 1 - T(\omega) - R(\omega)$ spectra for a thin InSb film and the proposed structure. In Panel~a, the spectra $T(\omega)$ and $R(\omega)$ are shown, and it is evident that the spectra of the bare film reveal behavior related to the beginning of Fabry-Perot oscillation. For the structure spectra, one can observe a resonant extremum related to the excitation of the local plasmons in the grating. In Panel~b, the comparison of the absorption spectra for the substrate and the proposed structure is shown. Both bare film spectra and the structure spectra are almost coincident for the non-resonant case (frequency range corresponding to the photon energies $<0.25\ eV$). Such a small difference emphasizes that the absorption in the grating is negligible under such conditions, and the prominent peak in the structure absorption spectrum is related mainly to the resonant plasmon absorption in the InSb film, but not the usual ohmic losses in the grating. It is worth noting that the magnitude of the effect is large enough: the plasmonic resonant absorption peak at value $~0.31$, which is bigger by an order of magnitude than absorption in the bare substrate, which peaks at approximately $0.025$. Although the full structure experiences more than a tenfold increase in absorption at the plasmon resonance, only the absorption in the InSb substrate is useful. In contrast, the absorption in the grating is considered parasitic.
To address this issue, we analyzed and compared the partial absorption in each component of the structure. A partial absorption within the InSb film can be calculated directly from the near-field distribution.
The local field $E(\textbf{r}, \omega)$ and the local time-averaged absorption $A(\textbf{r}, \omega)$ are related in the following way:
\begin{equation}\label{Absorption}
  A(\mathbf{ r}, \omega) = \frac{\omega}{c} \IM \varepsilon(\mathbf{ r}, \omega)|E(\mathbf{ r}, \omega)|^2
\end{equation}
where $\omega$ is a frequency of the incident wave, $c$ - is speed of light,  $\IM \varepsilon(\mathbf{ r}, \omega)$ is an imaginary part of the InSb dielectric function.

\begin{figure}[h]
  \centering
  \includegraphics[width=0.8\textwidth]{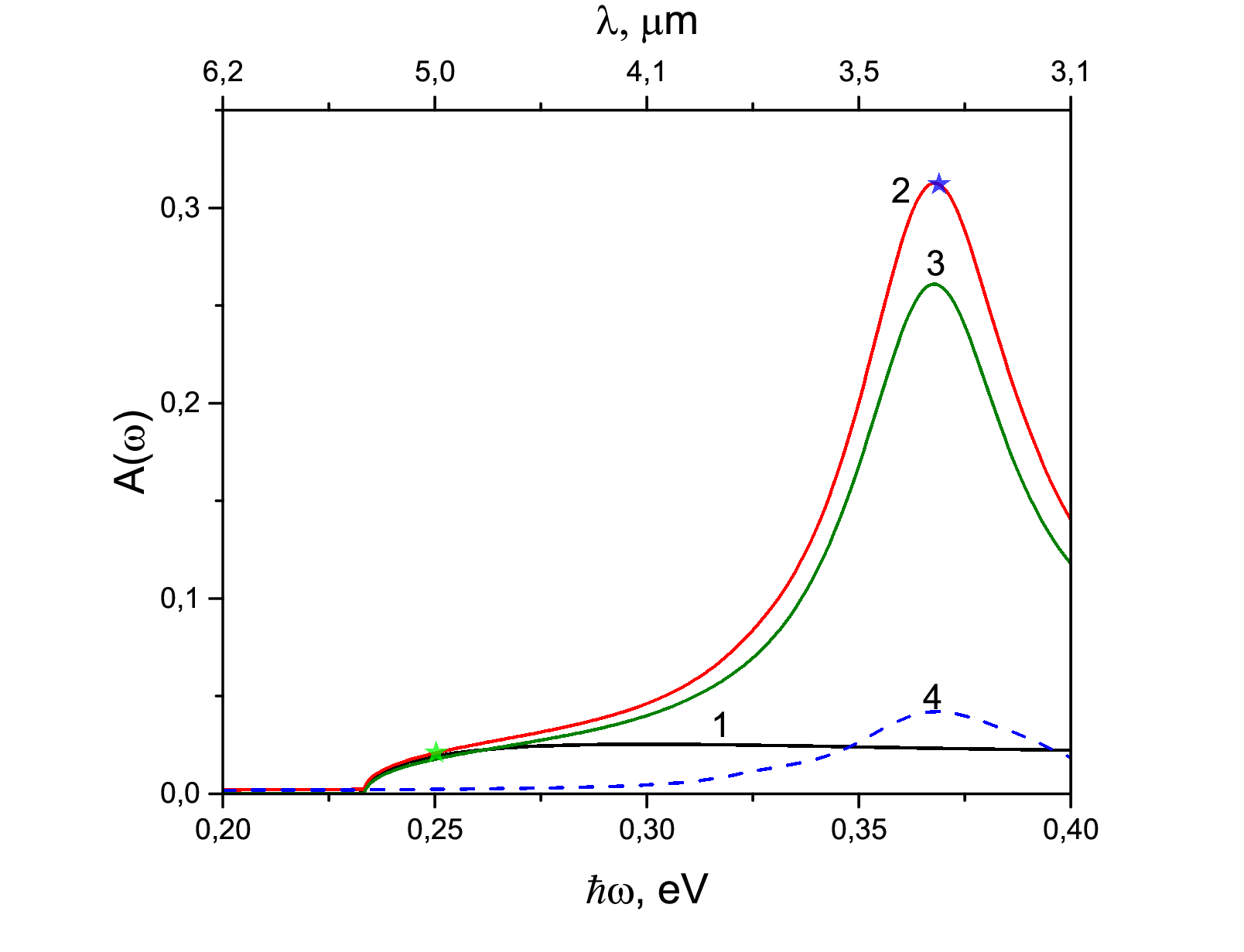}\\
  \caption{The spectra of absorption  of the structure and its components in comparison to the spectra of a bare substrate: 1 - the bare substrate, 2 - the full structure absorption, 3 - the partial absorption in the InSb layer, 4 -  the partial absorption in the metal grating. The parameters are the same as for Fig.~\ref{Fig2}.}
  \label{Fig3}
\end{figure}

 Fig.~\ref{Fig3} shows such an analysis. The peak value of useful absorption in the InSb plate is $~0.26$ (Curve 3). This is slightly smaller than the total absorption $~0.31$ (Curve 2) because the grating absorption is $~0.05$ (Curve 4) at the same frequency. Therefore, the grating absorption is much smaller than the useful absorption. Despite some level of parasitic absorption, we still have almost a tenfold increase in absorption. This result indicates excellent efficiency of the plasmonic structure. Moreover, we anticipate the possibility of further optimization, which could be achieved by using different materials for the grating or substrate, adjusting geometric parameters, or replacing the 1D-periodic grating with a 2D-periodic grating featuring rectangular shapes.

\begin{figure}[h]
  \centering
  \includegraphics[width=0.45\textwidth]{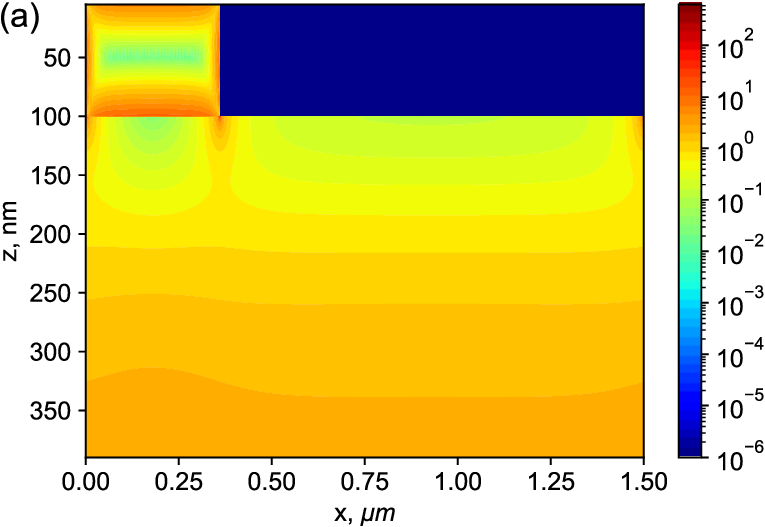}
  \includegraphics[width=0.45\textwidth]{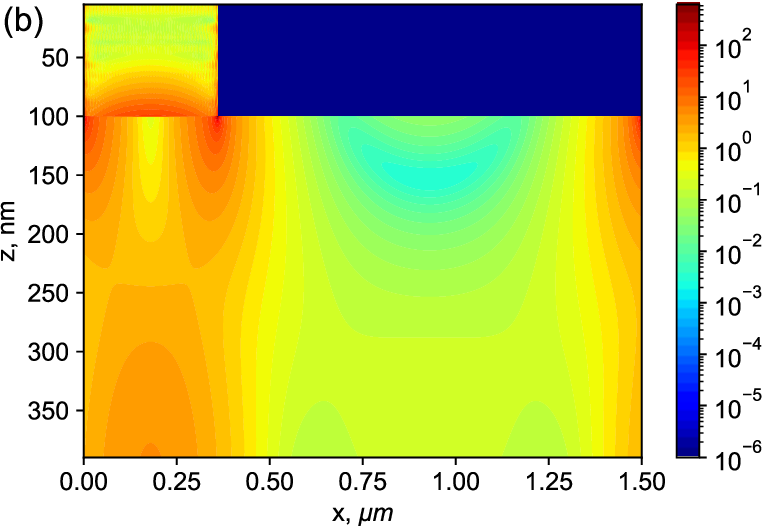}\\
  \caption{Panel a: the spatial distribution of the time-averaged absorption density for the non-resonant ($\hbar\omega = 0.25$ eV) and  Panel b: the resonant case ($\hbar\omega = 0.368$, which corresponds to the maximum of the absorption spectra).
   }\label{Fig4}
\end{figure}

Another feature of plasmonic resonance is the significant change in the local field, which is reflected in the local absorption picture (\ref{Absorption}). A strong redistribution of local absorption can be observed in its contour maps in Fig.~\ref{Fig4}. Panel (a) shows the spatial distribution for a non-resonant case. For this case, there is an almost uniform distribution along the x-axis, except for the small ``hot'' spots around the grating edges. In contrast to this, in the resonant case, a highly non-uniform distribution along the x-axis is observed. Also, the substantial expansion of the ``hot" areas reflects the resonant enhancement of the grating near field.
This redistribution has potential applications in designing efficient photodetectors. For instance, it may be possible to create areas of enhanced absorption in the proximity of the $p-n$-junction by tuning the geometric parameters of the structure.

\begin{figure}[h]
  \centering
  \includegraphics[width=0.45\textwidth]{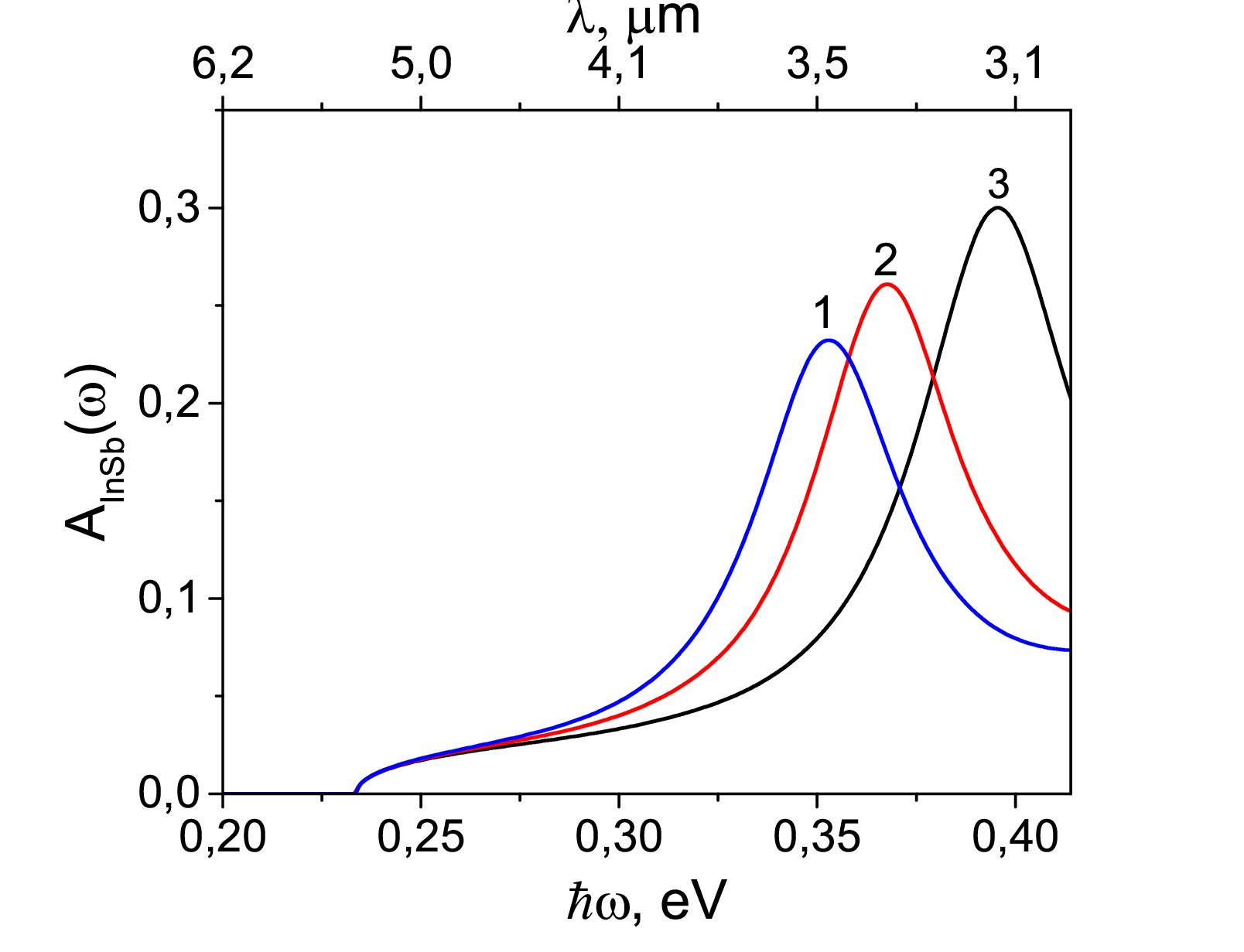}
  \includegraphics[width=0.45\textwidth]{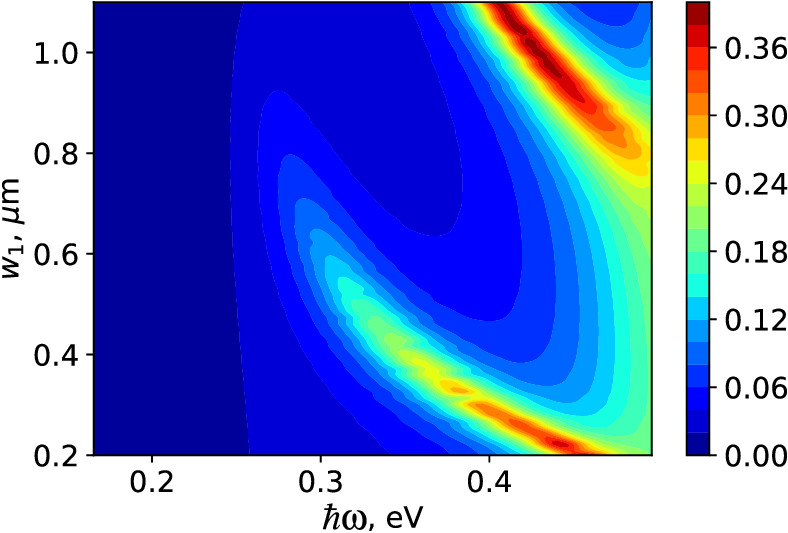}\\
  \caption{Panel a: The dependence of the spectra of absorption for the structures with different grating strip widths: 1, 2, 3, which correspond to 0.4, 0.36, and 0.3 $\mu m$, respectively. The other parameters are the same as for Fig.~\ref{Fig3}. Panel b: the contour mapping of the partial absorption spectrum depending on the strip width.}\label{Fig5}
\end{figure}
Unlike the frequency of the volume (3D) plasmon resonance, which depends solely on the intrinsic physical properties of the material, the characteristics of the local plasmon resonance exhibit a strong dependence on the geometric parameters of the structure as well.
Due to the grating plasmon resonance dependence on the geometrical parameters of the structure, it is possible to manipulate both the frequency and intensity of the absorption peaks.
Fig.~\ref{Fig5}(a) illustrates the dependence of plasmon resonance on the grating strip width $w$. This relationship is inverse: as $w$ increases, the plasma frequency decreases, and vice versa. For instance in Fig.~\ref{Fig5}(a), it was shown that the $\hbar\omega_p$ increases from $0.35$  at $w = 0.4$  (Curve 1) to $0.4$  at $w = 0.3$  (Curve 3). Although there is no rigorous theory to explain this dependence, simple limiting cases of grating geometry allow analytical estimation of the plasmon resonance frequency $\omega_p$. The theory of the grating resonance for a 2D-grating with the semielliptic profile of conductivity $\sigma_0^{2D}\sqrt{1-(2x/w)^2}$ was developed by Mikhailov \cite{MikhPRB98, Mikh1999Ideas}. In such a case, a simple expression for $\omega_p$ exists.
 \begin{equation}\label{omeega_p}
\omega_{p}^2 =\frac{16\pi n^{2D}e^2}{m^{*}\varepsilon_{eff} w}\left(1 - \frac{1}{24}\left[\frac{\pi w}{a}\right]^2 - \frac{1}{960} \left[\frac{\pi w}{a}\right]^4 \right)
 \end{equation}
 where $n^{2D} = n^{3D}d_g$ is an effective two-dimensional electron concentration, $m^{*},\ e$ is an electron effective mass and elementary charge in the Gauss units,  $\varepsilon_{eff}$ is an effective dielectric permittivity of the surrounding media. For the case of the grating between two uniform half-spaces with dielectric permittivities $\varepsilon_1$ and $\varepsilon_2$, it is simply  $\varepsilon_{eff} = (\varepsilon_1+\varepsilon_2)$. For the case of a grating on a substrate, $\varepsilon_{eff}$ has a cumbersome expression, but an expression for two half-spaces can be a good approximation for a rough estimate of $\varepsilon_{eff}$ if the $a/(2\pi)>>D_s$ is satisfied.
 If the grating thickness is less than the skin layer depth, this expression remains valid to some extent for gratings with a rectangular profile.
 For thick gratings, this expression does not allow direct estimation of $\omega_p$, but gives a qualitative explanation of the dependence $\omega_p$ on the grating strip width.

A more systematic approach to analyzing the plasmonic properties of such a structure involves contour mapping of the spectra with respect to the strip width $w$, which is shown in Fig.~\ref{Fig5}. Mikhailov's theory predicts only single-mode resonance existence\cite{MikhPRB98}. In contrast to this case, two distinct areas of plasmon resonance are observed in Fig.~\ref{Fig5}, corresponding to different modes. At the same frequencies, the second-order modes are excited within wider strips and exhibit greater intensity than the first-order modes. Although this fact is unexpected, it can be clearly explained by considering the characteristics of the near field generated at the edges of the strips. These fields intensify at larger filling factors $w/a$ \cite{My2012Grating, My2014Grat2D}, which may contribute to the enhancement of plasmon resonance intensity.

\section{Conclusion}

We propose a plasmonic structure consisting of an Indium Antimonide (InSb) substrate with a gold grating to enhance the absorption of infrared radiation in an ultrathin InSb film, enabling the potential development of highly sensitive multi-color detectors. This enhancement arises from the resonant features of the structure and the redistribution of the local field. The comparison of absorption spectra for the thin InSb substrate and the proposed structure reveals a more than ten-fold increase in absorption at the plasmonic resonance. Despite some parasitic absorption in the grating, there is still a significant enhancement in useful absorption, indicating the excellent efficiency of the proposed plasmonic structure. Future optimizations could involve using different materials for the grating or substrate, adjusting geometric parameters, or replacing the 1D-periodic grating with a 2D-periodic grating featuring rectangular shapes.
Building upon the proposed plasmonic structure, further research could explore the impact of various geometric parameters, such as grating periodicity and depth, on the absorption enhancement. By optimizing these parameters, it may be possible to tailor the resonant absorption to the different spectral ranges related to the specific applications, thereby improving the performance of InSb-based infrared radiation detectors. Additionally, the investigation of alternative materials for the grating or substrate could lead to the discovery of even more efficient plasmonic structures. Finally, studying the integration of the proposed structure with existing photodetector technologies, as well as the development of fabrication techniques to manufacture these enhanced devices could contribute to the creation of next-generation infrared detectors with improved sensitivity and multi-color detection capabilities.

\section{Acknowledgment}

The authors express their gratitude to the National Academy of Sciences of Ukraine for support within the framework of the project 0125U000799 ``New physical principles and technologies for the development of the element base of modern infrared photoelectronics''
and to the National Research Foundation of Ukraine, project  2025.06/0077  ``Technologies for forming passivating coatings for IR photodetectors to increase their detection capability and reliability''.

\bibliographystyle{unsrt}
\bibliography{sources}
\end{document}